\documentclass[%
 reprint,
 amsmath,amssymb,
 aps,
]{revtex4-2}

\usepackage{amssymb}
\usepackage{physics}
\usepackage{graphicx}
\usepackage{amsmath}
\usepackage{amsthm}
\usepackage{amsfonts}
\usepackage[T1]{fontenc}
\usepackage{array}
\usepackage{multirow}
\usepackage{color}
\usepackage{esint}
\usepackage{bm}
\usepackage{color}
\usepackage{bbm}
\usepackage{hyperref}
\usepackage{babel}

\usepackage{hyperref}
\usepackage[normalem]{ulem}
\hypersetup
{	colorlinks,%
	citecolor=green,%
	linkcolor=red,%
	urlcolor=blue%
}
\allowdisplaybreaks[4]

\begin{document}

\preprint{APS/123-QED}

\title{Exploring Unique Characteristics in Stark Many-Body Localization}

\author{Chung Po, CHING$^1$}
\email{cpching4-c@my.cityu.edu.hk}
\affiliation{$^1$Department of Physics, City University of Hong Kong, Kowloon, Hong Kong SAR}
\date{\today}
             \date{\today}

\begin{abstract}
Stark Many-Body Localization (MBL) is a phenomenon observed in quantum systems in the absence of disorder, where the presence of a linear potential, known as the Stark field, causes the localization. Our study aims to provide novel insight into the properties of Stark MBL and to discover unique entanglement characteristics specific to this phenomenon. The phase diagram analysis reveals different behavior with varying interaction strengths.  Furthermore, we highlight the influence of domain wall structures on the breakdown of the entanglement entropy of the system. Moreover, the investigation of Out-of-Time-Ordered Correlator (OTOC) behavior demonstrates distinct responses to interactions based on domain wall configurations. Our findings contribute to a better understanding of Stark MBL and offer valuable insights into the entanglement properties of systems subjected to Stark potentials.
\end{abstract}


\maketitle

\section{Introduction}
Many-body localization (MBL) has gained significant attention as a phenomenon observed in disordered interacting quantum systems, where particle localization persists even at high energy densities. MBL extends the concept of Anderson localization, which describes the localization of non-interacting particles in a disordered potential, to interacting quantum systems. Disorder in these systems can disrupt thermalization and give rise to localized states. Experimental investigations utilizing cold-atomic setups \cite{MBL_cold_atom_1,MBL_cold_atom_2,MBL_cold_atom_3,MBL_cold_atom_4,MBL_cold_atom_5}, trapped ions \cite{MBL_trap_ion_1,MBL_trap_ion_2,MBL_trap_ion_3,MBL_trap_ion_4,MBL_trap_ion_5}, and solid-state devices \cite{MBL_ssd_1,MBL_ssd_2,MBL_ssd_3} have been instrumental in the study of MBL.

The Stark field is able to localize non-interacting systems without disorders, a phenomenon dubbed Wannier-Stark localization. Recent research has focused on the interplay between MBL and external fields, particularly in the context of Stark MBL. The introduction of the linear potential in Stark MBL presents a novel mechanism in localization. Experimental studies exploring Stark MBL have been carried out in various systems, including ultracold atoms in optical lattices \cite{SMBL_ol_1,SMBL_ol_2,SMBL_ol_3,SMBL_ol_4} and semiconductor devices \cite{SMBL_scd_1,SMBL_scd_2}. Stark MBL provides many novel and interested localization mechanism.

Our study aims to provide novel insight into the properties of Stark MBL. We discover unique entanglement characteristics specific to Stark MBL, which expand our understanding of this phenomenon. The phase diagram analysis reveals different regimes, delineating the system's behavior with varying interaction strengths. We highlight the influence of domain-wall structures on the breakdown of the ETH and entanglement entropy, showcasing their impact on system dynamics. Moreover, the investigation of Out-of-Time-Ordered Correlator (OTOC) behavior demonstrates distinct responses to interactions based on domain wall configurations. These findings contribute to a better understanding of Stark MBL and offer valuable insights into the entanglement properties of systems subjected to Stark potentials, advancing our comprehension of Stark MBL.

\section{Model}
We consider a one-dimensional spinless fermion chain of length \textit{L} with open boundary condition. The Hamiltonian of our model is as follows: 
\begin{equation}
\begin{aligned}
	\hat{H} = & J\sum_{j=0}^{L-2} \left(c_{j}^{\dagger}c_{j+1}+h.c.\right)
+ \sum_{j=0}^{L-1} h_j\left(n_{j}-\frac{1}{2}\right)
\\ & + U\sum_{j=0}^{L-2}\left(n_{j}-\frac{1}{2}\right)\left(n_{j+1}-\frac{1}{2}\right).
\end{aligned}
\end{equation}
Here,  $c_{j}^{\dagger}$ ($c_{j}$) are the creation (annihilation) operators for fermions on the lattice site \textit{j}, and $n_j=c_{j}^{\dagger}c_{j}$ is the associated particle number operator. \textit{J} represents the strength of the nearest-neighbor hopping, while \textit{U} represents the strength of the nearest-neighbor interaction. The on-site potential for Stark model is represented as follows:
\begin{equation}
\begin{aligned}
	h_j=-{\gamma}j+{\alpha}\left(\frac{j}{L-1}\right)^2.
\end{aligned}
\end{equation}
The term $-{\gamma}j$ provides a linear potential with tilt ${\gamma}$, and ${\alpha}$ introduces a weak harmonic trap.

\section{Phase Diagram}
\begin{figure}[ht]
\centering
\includegraphics[width=1\columnwidth]{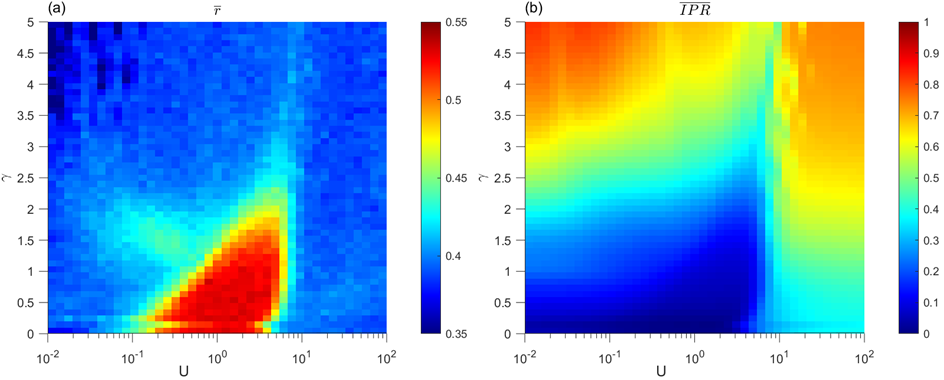}
\caption{\label{Fig: phase diagram}Sketch of the phase diagram for Stark MBL. Mean gap ratio (a) and inverse participation ratio (b) as a function of the strength of tilt and interaction.}
\end{figure}

We evaluate the MBL phase diagrams using the mean gap ratio and the many-body inverse participation ratio (IPR) with the size of the system $L=16$ at half-filling with the harmonic trap ${\alpha}=1$ as shown in Fig.~\ref{Fig: phase diagram}.   

Level statistics is a basic but powerful tool that distinguishes the thermal and Stark MBL phases \cite{LS_MBL_1,LS_MBL_2,LS_MBL_3, LS_MBL_5}. The gap ratio is defined as 
\begin{equation}
\begin{aligned}
r_i={\rm min}\left\{ \frac{{\delta}E_i}{{\delta}E_{i+1}},\frac{{\delta}E_{i+1}}{{\delta}E_i} \right\},
\end{aligned}
\end{equation}
where ${\delta}E_i=E_{i+1}-E_i$ and $E_i$ is $i^{th}$ eigenenergy. We then averaged all the eigenstates to obtain the mean gap ratio $\overline{r}$. The spectrum follows the Gaussian orthogonal ensemble (GOE) with $\overline{r}=0.530$ in the thermal phase and the Poisson distribution with $\overline{r}=0.386$ in the MBL phase. The Stark MBL can also be identified through the localization of the eigenstate in real space, qualified by the many-body IPR \cite{IPR_MBL_1}, which is defined as

\begin{equation}
\begin{aligned}
IPR^{(\epsilon)}=\frac{1}{1-\nu}\left(\frac{1}{{\nu}L}\sum_{j=1}^{L}\left|u_j^{\epsilon}\right|^{2}-\nu\right) ,
\end{aligned}
\end{equation}
where $u_j^{\epsilon}=\left<{\psi}^{\epsilon}\right|n_j\left|{\psi}^{\epsilon}\right>$, projected particle number of eigenstate $\left|{\psi}^{\epsilon}\right>$ with eigenenergy $\epsilon$, and $\nu$ is filling factor.  The thermal (localized) states are characterized by $IPR{\rightarrow}\left({\rightarrow}1\right)$. Then we averaged over all the eigenstates to obtain  $\overline{IPR}$. 

For weak interactions, the system is dominated by single-particle properties. It indicates the non-interacting Stark localization transition at $\gamma = 4/L$, with the whole spectrum localized (extended) for $\gamma>4/L$ $\left(\gamma<4/L\right)$. In this regime, the system is weak to thermalize, making it difficult to observe a clear transition in localization through level statistics. As the interaction strength increases to an intermediate regime, the thermal phase extends to $\gamma\sim3.2$. The system consistently exhibits Stark many-body localization (MBL) when the tilt is above this value. All diagrams in this regime indicate the emergence of an ergodic phase. However, as the system enters the strongly interacting regime, it becomes almost localized due to Hilbert space fragmentation \cite{HSF_1,HSF_2,HSF_3}. This results in the isolation of distinct sectors within the Hilbert space, which in turn leads to the persistence of non-thermal behavior in the system.

\section{Entanglement Entropy}
\begin{figure}[ht]
\centering
\includegraphics[width=\columnwidth]{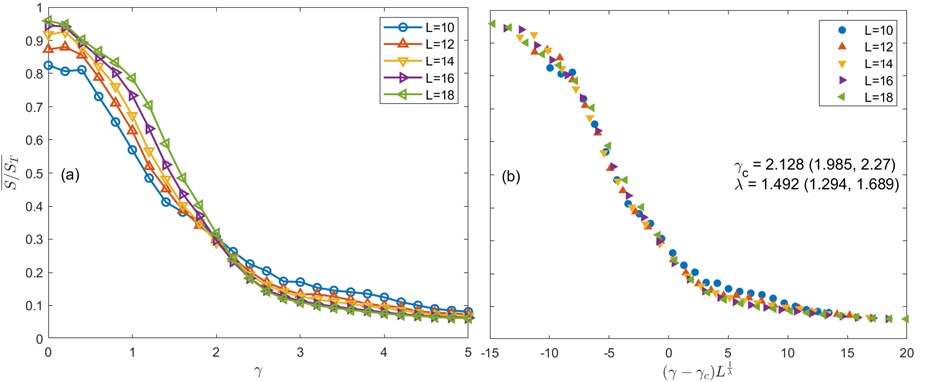}
\caption{\label{Fig: Entropy}Average half-chain entanglement entropy divided by the Page value $S_T$ (a) and finite-size critical scaling collapse (b) for the Stark model.}
\end{figure}

To further the Stark MBL transition, we perform a finite-size scaling analysis on the half-chain entanglement entropy to investigate the Stark MBL transition across various system sizes at half filling with the harmonic trap, ${\alpha}=1$, and the nearest-neighbor interaction, $U=1$, as shown in Fig.~\ref{Fig: Entropy}. 

\begin{figure*}[ht]
\centering
\includegraphics[width=1\textwidth]{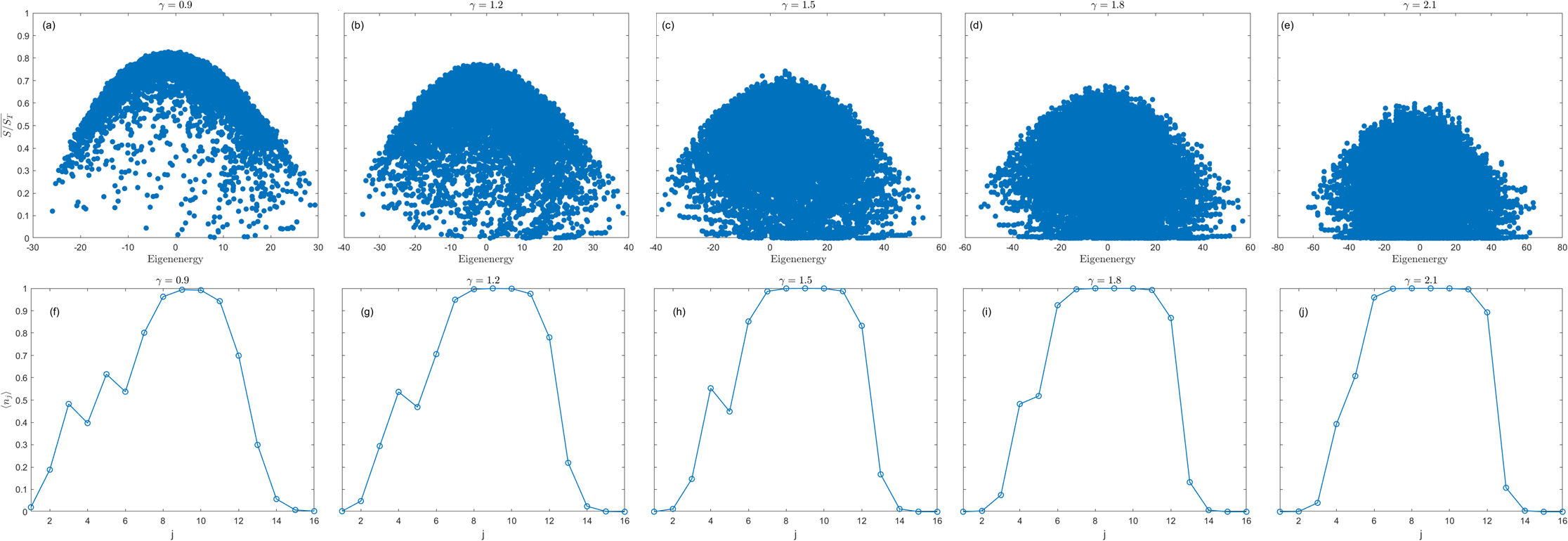}
\caption{\label{Fig: En_State}Upper panel (a to e) show the entanglement entropy of eigenstates with different tilt. The lower panel (f to h) shows the expectation of the particle number corresponding to one of the lowest entanglement entropy eigenstates with different tilt.
}
\end{figure*}

The half-chain entanglement entropy is defined as
\begin{equation}
\begin{aligned}
S=-{\rm Tr}\left({\rho}_{\rm A}{\rm ln}{\rho}_{\rm A}\right),
\end{aligned}
\end{equation}
where ${\rho}_{\rm A}={\rm Tr_{B}}\left(\left|{\psi}\right>\left<{\psi}\right|\right)$ is the reduce density matrix for subsystem A, which corresponds to the left half of the chain, by tracing over another subsystem B, representing the right half of the chain, in a composite system with eigenstate $\left|{\psi}\right>$.The results are averaged over the middle quarter of the eigenstates and then divided by the Page value \cite{PageV_1}, $S_T=0.5\left[L \ln(2)-1\right]$. We then perform the scaling data collapse $\overline{S/S_T}$ by fitting it to the form $f\left[\left({\gamma}-{\gamma}_c\right)L^{\frac{1}{\lambda}}\right]$, where ${\gamma}_c$ is the critical tilt and ${\lambda}$ is the scaling exponent.  Transition of the entanglement from the volume law $\overline{S/S_T} {\rightarrow}1$  in the thermal phase to the area law $\overline{S/S_T} {\rightarrow}0$  in the MBL phase. The critical tilt estimated by the data of entanglement entropy is ${\gamma}_c\sim 2.13$, and the scaling exponent is $\lambda\sim1.49$.  Compared to disorder-driven MBL, we have observed significant differences in the Stark model. Specifically, the behavior of the half-chain entanglement entropy deviates from the expected step function pattern, as observed in disorder-driven MBL \cite{Dis_EE_1,Dis_EE_2}. Instead, we find that it exhibits a decrease when a finite tilt is applied. We attribute this feature to the proliferation of the domain wall state, which will be elaborated on in the next section. Furthermore, the scaling exponent in the Stark model is also larger than that found in disorder-driven MBL.

\section{Domain Wall}

In this section, we discuss the properties and formation of domain wall state. Domain walls form as boundaries or regions of contrasting properties within the system, arising from the interplay between the Stark potential and interactions. Their formation results in localization, where the occupied properties on either side of the wall differ. The double-domain wall can be constructed by combining two states of the single-domain wall. One state corresponds to the ground state, where all sites are occupied on the right-hand side ($...\circ\circ\circ\bullet\bullet\bullet...$), while the other state represents the highest excited state, with all sites occupied on the opposite side ($...\bullet\bullet\bullet\circ\circ\circ...$).  The double domain wall state ($...\circ\circ\circ\bullet\bullet\bullet...\bullet\bullet\bullet\circ\circ\circ...$) can be formed by combining these two states, where they have the same occupied properties at their ends. It is important to note that the distance between each domain wall is required to be larger than a correlation length, $\xi$.  Therefore, the system can be divided into at most N pieces, $N<L/{\xi}$. The number of eigenstates with a domain wall structure, comprising $2^{L/\xi}$ states, can be approached using this construction. When the interaction is applied to the domain wall formed by the ground state, it acts as a repulsion, resulting in the formation of kinks and making it broader than the other side. In contrast, when the interaction is applied to the domain wall formed by the highest excited state, it acts as an attraction that creates a sharp boundary.

We first investigate through exact diagonalization for the small system with size $L=16$ at half-filling, with a harmonic trap $\alpha=1$. We can observe that a state with low entanglement entropy exhibits a domain-wall structure as shown in Fig.~\ref{Fig: En_State}. The domain-wall states account only for an infinitesimal portion for the whole spectrum in the thermalization limit. Hence, there are rare states embedded in the thermal spectrum, breaking the Eigenstate Thermalization Hypothesis (ETH) in a strong sense \cite{DW_S_1}. Interestingly, as the tilt increases, we find that more states generate domain-wall structures. Furthermore, the breakdown of ETH associated with the domain wall has a direct impact on the entanglement entropy of the system. Specifically, when a finite tilt is applied, we observe a drop or decrease in the entanglement entropy. This observation highlights the influence of the domain wall and the breakdown of ETH on the entanglement entropy.

\begin{figure}[ht]
\centering
\includegraphics[width=\columnwidth]{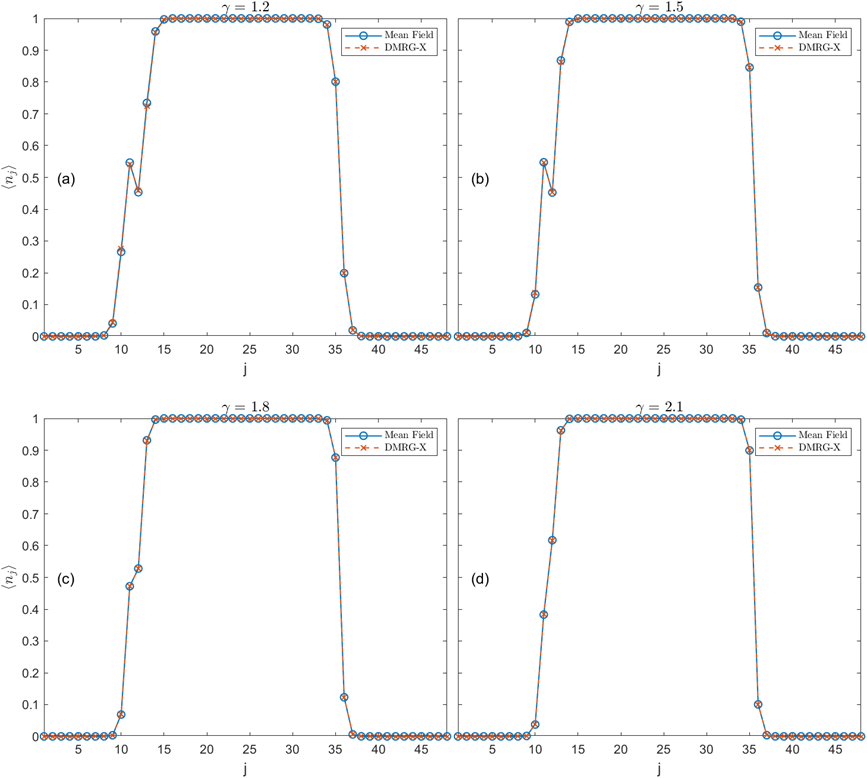}
\caption{\label{Fig: Mean field}The expectation of the particle number in the mean field with double domain wall states and their corresponding eigenstates derived by DMRG-X with different tilt. For the DMRG-X algorithm, the energy uncertainty is $\left<H^2\right>-\left<H\right>^2=1.8\times{10^{-9}}$ for $\gamma=1.2$, $9.1\times10^{-12}$ for $\gamma=1.5$, $2.0\times10^{-11}$ for $\gamma=1.8$, and $-1.5\times10^{-11}$ for $\gamma=2.1$. Here, we take $U=1$. }
\end{figure}

Additionally, we employ mean-field methods to study domain wall structures for a large system with size $L=48$ at half-filling, without a harmonic trap. We use the double domain wall state as an initial guess for our investigation. These domain-wall structures can also be observed in large systems. The system exhibits a preference for domain wall formation to minimize energy. To compare the Mean Field and DMRG-X, we obtain a high overlap $\left(\geq 0.9998\right)$, as shown in Fig.~\ref{Fig: OTOC non interact}. Additionally, the initial position of the domain walls also allows for shifting within the system while maintaining the fixed distance between them. 

\section{Out-of-time-ordered correlator}

We evaluated OTOC in many body systems with a system size $L=24$ in half-filling without harmonic trap and with different tilts introduced into the system. The OTOC is a quantity used to probe quantum chaos and thermalization \cite{OTOC_1,OTOC_2,OTOC_3}. The OTOC with a pure state $\left|{\psi}_{0}\right>$ is defined as  
\begin{equation}
\begin{aligned}
F\left(t\right)=\left<{\psi}_{0}\right|\hat{W}\left(t\right)\hat{V}\hat{W}^{\dagger}\left(t\right)\hat{V}^{\dagger}\left|{\psi}_{0}\right>.
\end{aligned}
\end{equation}
In our work, we take the single domain wall states as initial states and $\hat{W}={{\sigma}^{z}_{12}}$ , is set on the domain wall, and $\hat{V}={{\sigma}^{z}_{j}}$. The equation can be simplified to $F\left(t\right)=\left<{\psi}_{z}\left(t\right)\right|{{\sigma}^{z}_{j}}\left|{\psi}_{z}\left(t\right)\right>\left<{\psi}_{0}\right|{{\sigma}^{z}_{j}}\left|{\psi}_{0}\right>$, and $\left|{\psi}_{z}\left(t\right)\right>={{\sigma}^{z}_{12}}\left|{\psi}_{0}\right>$. 

\begin{figure}[ht]
\centering
\includegraphics[width=\columnwidth]{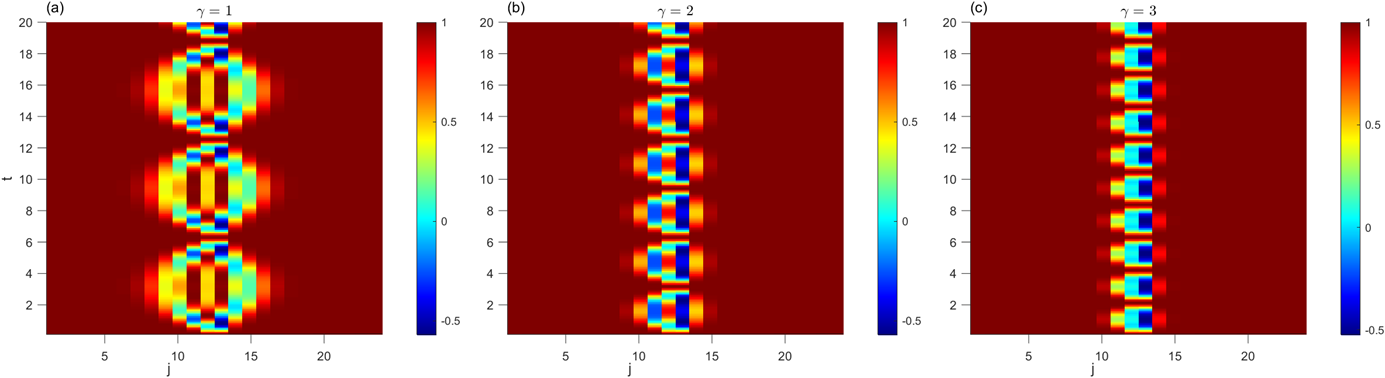}
\caption{\label{Fig: OTOC non interact}OTOC using a single domain wall state under the non-interacting limit with different tilt. The oscillation frequency and extended distance decrease properties when the tilt increases.}
\end{figure}
For non-interacting limit, we observe that the system can always be localized with an oscillation pattern as shown in Fig.~\ref{Fig: OTOC non interact}.  The oscillation period is determined to be $2\pi/\gamma$, and the width of the oscillation corresponds to the width of the domain wall. Initially, it was located at the center of the chain spread in both directions. After reaching a certain distance from its initial position, it contracts back towards its original location. Importantly, we observed that regardless of the introduced tilts, localization consistently occurred when the system size exceeded the width of the oscillation. When the tilt is increased, both the oscillation frequency and the extended distance proportional decrease. In other words, as long as the system sizes were larger than the extent of the oscillation, it remained localized.
\begin{figure}[ht]
\centering
\includegraphics[width=\columnwidth]{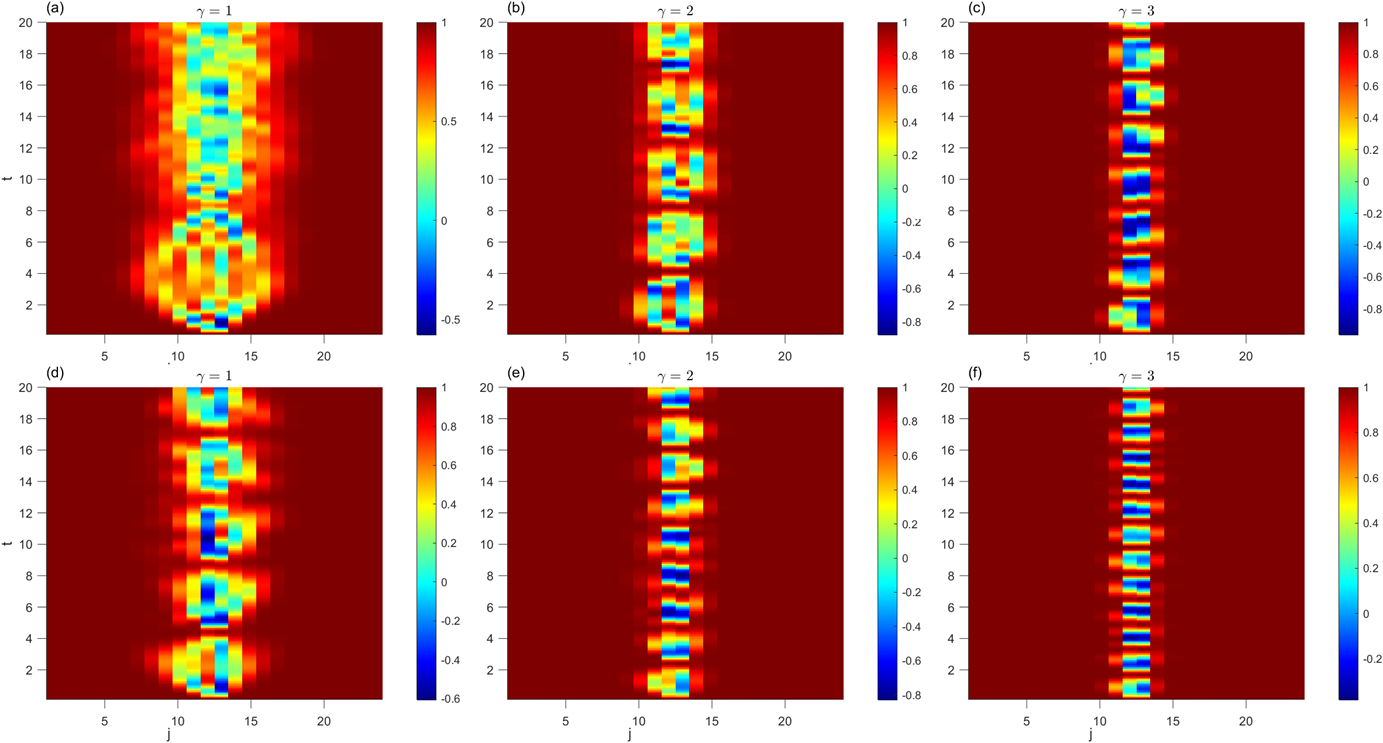}
\caption{\label{Fig: OTOC_Interact}OTOC using a single domain-wall state with an interaction strength of $U=1$. Upper panel (a to c) show the result with $\left|...\circ\circ\circ\bullet\bullet\bullet...\right>$  as initial state. Lower panel (d to f) show the result with  as initial state. $\left|...\bullet\bullet\bullet\circ\circ\circ...\right>$}
\end{figure}
When interactions are introduced with $U=1$, distinct results emerge along and presence of chaos as shown in Fig.~\ref{Fig:  OTOC_Interact}. In the case of the domain wall state associated with the formation of kink, the oscillation frequency decreases when the interaction acts as repulsion. Moreover, the oscillation pattern does not manifest when the tilt is smaller than the critical value. We also observe the presence of chaos in this scenario. Conversely, for the domain wall state associated with the formation of sharp boundary, the oscillation frequency increases due to attractive interactions. Remarkably, this state still exhibits the oscillation pattern even when the tilt is lower than the critical value.

\section{DISCUSSION AND CONCLUSION}

Our study has provided valuable insights into the distinct properties of Stark MBL. We have discovered important differences in the entanglement characteristics of Stark MBL, shedding light on its unique nature. By examining the phase diagram, we summarize the key changes that occur as the interaction strengths vary, allowing us to identify different regimes within the system. In the weakly interacting regime, single-particle properties dominate, whereas in the intermediate regime, Stark MBL and an emergent ergodic phase are consistently observed. Conversely, the strong regime exhibits Hilbert-space fragmentation.

Furthermore, we have emphasized the influence of domain wall structures and their impact on the breakdown of the ETH and the entanglement entropy. Through our analysis, we have revealed that the formation of domain walls is influenced by the type of interaction present, with repulsive interactions leading to the formation of kinks and attractive interactions resulting in sharp boundary. These domain wall structures play a crucial role in the entanglement properties of the system.

Additionally, we have investigated the behavior of the OTOC in many-body systems, providing insights into the system's response to interactions based on the domain wall configuration. Our findings demonstrate distinct responses, with the OTOC oscillation frequency decreasing with repulsive interactions in the single domain wall state with kink, while increasing with attractive interactions in the single domain wall state with sharp boundary. These results highlight the intricate relationship between interactions, domain walls, and the dynamics of the system.

Overall, our study contributes significantly to the understanding of Stark MBL. By elucidating its distinct entanglement characteristics compared to disorder-driven MBL, we have expanded our knowledge of the underlying mechanisms at play. The insights gained from our analysis of the phase diagram, domain-wall structures, and OTOC behavior provide a comprehensive understanding of Stark MBL and its implications for the entanglement properties of systems. This knowledge furthers our comprehension of Stark MBL.

\bibliography{draft}

\end{document}